\documentclass[aps,prb,reprint,superscriptaddress,amsmath,amssymb,floatfix,twocolumn]{revtex4-1}
\usepackage{times}
\usepackage{graphicx}
\usepackage{subfigure}
\usepackage{color}

\begin{document}

\title{Slave spin theory of magnetic states in Hubbard model}

\author{Jiasheng Qian}
\affiliation{Department of Physics and Beijing Key Laboratory of Opto-electronic Functional Materials \& Micro-nano Devices, Renmin University of China, Beijing 100872, China}

\author{Zhiguang Liao}
\affiliation{Department of Physics and Beijing Key Laboratory of Opto-electronic Functional Materials \& Micro-nano Devices, Renmin University of China, Beijing 100872, China}

\author{Rong Yu}
\email{rong.yu@ruc.edu.cn}
\affiliation{Department of Physics and Beijing Key Laboratory of Opto-electronic Functional Materials \& Micro-nano Devices, Renmin University of China, Beijing 100872, China}
\affiliation{Key Laboratory of Quantum State Construction and Manipulation (Ministry of Education), Renmin University of China, Beijing, 100872, China}

\begin{abstract}
Magnetic properties of Hubbard model have been studied extensively. A theoretical description of these states, however, is not straight forward within conventional mean-field approach due to the electron correlation effects. Here we provide a slave spin theory of the magnetic states in the single band Hubbard model. By introducing an additional decomposition of the on-site Coulomb interaction in the magnetic channel in both the slave spin and the fermionic spinon sectors, we show a magnetic solution with renormalized quasiparticle spectral weight can be stabilized at both half-filling and finite doping. By comparing the energies of antiferromagnetic and ferromagnetic states, we provide a ground-state phase diagram for generic electron filling. We further discuss possible generalization of the slave-spin method.
\end{abstract}

\maketitle


\section{Introduction}
One major theme of condensed matter physics is to study magnetism in correlated electron systems.
It has been recognized that many exotic phenomena, including unconventional supercoductivity, non-Fermi liquid, and electronic nematicity, are closely connected to magnetic ordering and fluctuations~\cite{Dagotto_RMP_1994, Lee_RMP_2006, Dai_RMP_2015, Si_NRM_2016, Hirschfeld_CRP_2016, Wang_Science_2011, Imada_RMP_1998, Stewart_RMP_2001, Fradkin_RMP_2015, PaschenSi_NRP_2021}. Therefore, investigating magnetic properties helps uncover the origin of these phenomena in strongly correlated systems.

A standard model for studying strong electron correlation effects in the Hubbard model~\cite{Arovas_ARCMP_2022}. It has been widely accepted that its single band and multiorbital versions can be used to describe many strong correlation effects in cuprate and iron-based superconductors. The phase diagram of Hubbard models contains rich magnetic states. As a well known example, a N\'{e}el antiferromagnetic (AFM) order is stabilized as the ground state of the single band Hubbard model on the two-dimensional (2D) square lattice with nearest neighbor (n.n.) hopping at half-filling. The origin of this AFM order can be understood from either weak or strong coupling limit. In the weak coupling limit, the AFM order driven by Fermi surface nesting, which makes the Fermi liquid unstable to a AFM spin density wave order. Alternatively, in the strong coupling picture, the AFM order is a natural result of the AFM exchange interaction between neighboring spins once the charge excitations are gapped out. However, it is still a mystery how the system crosses over from weak to strong coupling because the above two scenarios are not compatible and a unified theoretical description covering both limits is demanded. Actually the half-filling limit is special because a metal-to-Mott insulator transition takes place. But the situation is still not fully clear when the system is away from half-filling in the correlated metallic state, especially on how the magnetic order is established and how the bands are renormalized by the correlation effects.

One theoretical tool in studying Hubbard model is the slave-spin theory. Recently it has been widely used to study the metal to insulator transition and orbital selectivity in single band and multiorbital Hubbard models~\cite{Medici_PRB_2005, Medici_PRL_2009, Yu_PRB_2011, Medici_PRB_2011, Yu_PRB_2012, Yu_PRL_2013, Medici_PRL_2014, Fanfarillo_PRB_2015, Yu_PRB_2017, Komijani_PRB_2017, Yu_PRL_2018, Yang_CPB_2019, Pizarro_JPC_2019, Arribi_PRB_2021}. It has been shown that this method is able to describe the Mott transition at commensurate filling and in the metallic phase the obtained quasiparticle spectral weight $Z$ agrees with that in the Gutzwiller approximation~\cite{Medici_PRB_2005, Yu_PRB_2011, Yu_PRB_2012}. Compared to dynamical mean-field theory (DMFT)~\cite{Georges_RMP_1996, Kotliar_RMP_2006}, it is more efficient and can also capture the central physics in both noninteracting and strong coupling limits.

Despite its success, the slave spin theory is mostly used to study the properties of correlated systems in the paramagnetic (PM) phase. A slave-spin solution that can describe magnetic states at both half-filling and finite doping is, to the best of our knowledge, not established. The reason lies in the structure of the slave-spin formulation. In this theory, a slave quantum $S=1/2$ spin is introduced to carry the charge degree of freedom of electrons, and a fermionic spinon is used to carry their spin degree of freedom. At the mean-field level, the slave spin and spinon operators are decoupled. In a magnetic state, the magnetic order necessarily breaks the spin rotational symmetry. Therefore, to settle a magnetic ordered state in the slave-spin theory, the symmetry must be broken in both slave spin and spinon channels. But at mean-field level this is not straightforward given the decoupled nature of the slave spins and spinons. In the slave-spin theory, a Lagrangian multiplier is introduced to enforce the constraint of slave spin and spinon degrees of freedom to the physical sector. This Lagrangian multiplier appears in Hamiltonians of both channels. A na\"{\i}ve idea to obtain the magnetic solution is to make this multiplier to be different for different spin flavors. However, we show in below, based
on an energy argument, that this idea does not work. We then propose a new approach for the magnetic solution within the slave-spin formulation by introducing an additional decomposition of the on-site Coulomb interaction in the magnetic channel in both the slave spin and the fermionic spinon sectors. We check the validity of our method by studying AFM and ferromagnetic (FM) states in the single band Hubbard model at generic electron filling factor and provide a magnetic phase diagram of the system.

The remainder of the paper is organized as follows. In Sec.~\ref{Sec:Model}, we first introduce the single-band Hubbard model and outline the $U(1)$ slave-spin approach in the PM phase. Second, we provide reasons on why the na\"{\i}ve generalization by allowing the Lagrangian multipliers spin dependent fails to give a magnetic solution of the model. We then develop a new approach to get the magnetic solution in the slave-spin theory by introducing an additional decomposition of the on-site Coulomb interaction in the magnetic channel in both the slave spin and the fermionic spinon sectors. In Sec.~\ref{Sec:Results}, we show results of both FM and AFM solutions, and provide a ground-state phase diagram for generic electron filling. We further discuss effects of correlations and possible ways in generalizing the method in Sec.~\ref{Sec:Discussions}.

\section{Model and Method}
\label{Sec:Model}

\subsection{Single band Hubbard model and $U(1)$ slave-spin approach in the paramagnetic phase}

In this work, we consider a single band Hubbard model defined on a square lattice whose Hamiltonian takes the following form,
\begin{equation}
 \label{Eq:Ham_tot}
 H=\frac{1}{2}\sum_{ij\sigma} (t_{ij}-\mu\delta_{ij})
 d^\dagger_{i\sigma} d_{j\sigma} + \frac{U}{2} \sum_{i,\sigma}(n_{i\sigma}-\frac{1}{2}) (n_{i\bar{\sigma}}-\frac{1}{2}).
\end{equation}
where $d^\dagger_{i\sigma}$ creates an electron with spin $\sigma$ at site $i$, $n_{i\sigma}=d^\dagger_{i\sigma} d_{i\sigma}$, $\delta_{ij}$ is the Kronecker $\delta$ function, $\mu$ is the chemical potential, and $U$ is the
onsite Coulomb repulsion. $t_{ij}$ is the hopping parameter between site $i$ and $j$. Here we consider n.n. hopping $t_1$ and next n.n. hopping $t_2$.

We study this model by using a $U(1)$ slave-spin theory, which was introduced initially for studying the metal-insulator transition and orbital selectivity in the PM phase for multiorbital Hubbard models in Ref.~\onlinecite{Yu_PRB_2012}. Here we briefly summarize the approach to set the stage for our consideration of stabilizing magnetic solutions in the Hubbard model. For further details, we refer to Refs.~\onlinecite{Yu_PRB_2012} and \onlinecite{Yu_PRB_2017}.

In the $U(1)$ slave-spin formulation, we introduce a quantum $S=1/2$ spin operator whose XY component  ($S^{+}_{i\sigma}$) is used to represent the charge degree of freedom of the electron at each site $i$
and each spin flavor $\sigma$. We also introduce a fermionic ``spinon'' operator
($f^\dagger_{i\sigma}$) to carry the spin degree of freedom. The electron creation operator is then rewritten as
\begin{equation}
 \label{Eq:SSCreate} d^\dagger_{i\sigma} = S^+_{i\sigma} f^\dagger_{i\sigma}.
\end{equation}
To restrict the Hilbert space of the slave spin representation to the physical one, we enforce a local constraint,
\begin{equation}
 \label{Eq:constraint} S^z_{i\sigma} = f^\dagger_{i\sigma} f_{i\sigma} - \frac{1}{2}.
\end{equation}

This representation contains a $U(1)$ gauge redundancy corresponding to
$f^\dagger_{i\alpha\sigma}\rightarrow f^\dagger_{i\alpha\sigma} e^{-i\theta_{i\alpha\sigma}}$
and $S^+_{i\alpha\sigma}\rightarrow S^+_{i\alpha\sigma} e^{i\theta_{i\alpha\sigma}}$.
As a consequence, the slave spins carry the $U(1)$ charge. Note that the $U(1)$ gauge structure here is different from the $Z_2$ slave spin representation~\cite{Medici_PRB_2005,Nandkishore_PRB_2012}.

To obtain the correct quasiparticle spectral weight within a mean-field approach, we define a dressed operator, in a way similar to the standard slave-boson theory~\cite{KotliarRuckenstein}:
\begin{equation}
 \label{Eq:Zdagger} \hat{z}^\dagger_{i\sigma} = P^+_{i\sigma} S^{+}_{i\sigma} P^-_{i\sigma},
\end{equation}
where $P^\pm_{i\sigma}=1/\sqrt{1/2\pm S^z_{i\sigma} +\delta}$, and $\delta$ is an infinitesimal positive
number to regulate $P^\pm_{i\sigma}$.
Eq.~\eqref{Eq:SSCreate} now becomes
\begin{equation}\label{Eq:SBcreate}
d^\dagger_{i\sigma}=\hat{z}^\dagger_{i\sigma} f^\dagger_{i\sigma}.
\end{equation}
Consequently, the Hamiltonian, Eq.~\eqref{Eq:Ham_tot},
can then be effectively rewritten, up to a constant, as
\begin{eqnarray}
\label{Eq:HamSS} H &=& \frac{1}{2}\sum_{ij\sigma} t_{ij}
 \hat{z}^\dagger_{i\sigma} \hat{z}_{j\sigma} f^\dagger_{i\sigma} f_{j\sigma}
 -\mu \sum_{i\sigma}  f^\dagger_{i\sigma}
 f_{i\sigma}
 \nonumber\\
 && - \sum_{i\sigma} \lambda_{i\sigma}\left(f^\dagger_{i\sigma}
 f_{i\sigma}-S^z_{i\sigma}\right) + \frac{U}{2} \sum_i \left( \sum_{\sigma} S^z_{i\sigma} \right)^2.
\end{eqnarray}
Here, $\lambda_{i\sigma}$ is a Lagrange multiplier to enforce the constraint in Eq.~\eqref{Eq:constraint}.
The interaction Hamiltonian is rewritten by using slave-spin operators~\cite{Yu_PRB_2012}.
The quasiparticle spectral weight is defined as
\begin{equation}
\label{Eq:qpWeightZ}
Z_{i\sigma} = |\langle \hat{z}_{i\sigma} \rangle|^2.
\end{equation}

Note that Eq.~\eqref{Eq:HamSS} still includes strong interactions between slave spins. To further simplify the problem, we decompose the slave-spin and spinon operators at the mean-field level, treat the constraint on average, and make a single-site approximation for the slave spin operators by making use of the translational symmetry of the system. In this way, we can then drop the site index for slave spins and $\lambda_{i\sigma}$.
To get the correct quasiparticle spectral weight at the noninteracting limit ($U=0$), we further approximate that
\begin{equation}\label{Eq:zapprox}
\hat{z}^\dagger_{\sigma} \approx {\tilde{z}}^\dagger_{\sigma} + 2\langle {\tilde{z}}^\dagger_{\sigma} \rangle \eta_{\sigma} \left( S^z_{\sigma}+1/2-n^f_{\sigma} \right),
\end{equation}
where
\begin{equation}
 \tilde{z}^\dagger_{\sigma} = \langle P^+_{\sigma}\rangle S^{+}_{\sigma} \langle P^-_{\sigma} \rangle,
\end{equation}
\begin{equation}
 n^f_{\sigma}=\frac{1}{N} \sum_k \langle f^\dagger_{k\sigma} f_{k\sigma}\rangle,
\end{equation}
and
\begin{equation}
 \eta_{\sigma}= \frac{2n^f_{\sigma}-1}{4n^f_{\sigma}(1-n^f_{\sigma})}.
\end{equation}

Note that $\langle \hat{z}_{\sigma} \rangle = \langle \tilde{z}_{\sigma} \rangle$, and by making use of the constraint in Eq.~\eqref{Eq:constraint}, we see that the second term of Eq.~\eqref{Eq:zapprox} introduces an additional effective chemical potential term to the Hamiltonian
\begin{equation}\label{Eq:mutilde}
 \sum_{k\sigma} {\tilde{\mu}}_{\sigma}  (f^\dagger_{k\sigma} f_{k\sigma} - n^f_{\sigma}),
\end{equation}
where
\begin{equation}
\label{Eq:tilde-mu}
\tilde{\mu}_{\sigma} = 2\bar{\epsilon}_{\sigma} \eta_{\sigma},
\end{equation}
\begin{equation}
\label{Eq:tilde-bar-epsilon}
\bar{\epsilon}_{\sigma} = \sum_{\sigma^\prime}\left( Q^f \langle\tilde{z}_\sigma^\dagger\rangle \langle \tilde{z}_{\sigma^\prime} \rangle  + \rm{c.c.} \right),
\end{equation}
and
\begin{equation}
\label{Eq:Qf}
Q^f = \sum_{k\sigma}\epsilon_k\langle f^\dagger_{k\sigma}
f_{k\sigma}\rangle/2,
\end{equation}
with $\epsilon_k=\frac{1}{N}\sum_{ij\sigma} t_{ij} e^{ik(r_i-r_j)}$ the band dispersion.

Then the dynamical part of the mean-field Hamiltonians for the spinons and the slave spins are, respectively,
\begin{eqnarray}
 \label{Eq:Hfmf}
 H^{\mathrm{mf}}_f &=&  \sum_{k\sigma}\left[ \epsilon_{k} \langle \tilde{z}^\dagger_{\sigma} \rangle
  \langle \tilde{z}_{\sigma} \rangle + (\tilde{\mu}_\sigma-\lambda_\sigma-\mu)\right] f^\dagger_{k\sigma} f_{k\sigma}\\
 \nonumber \\
 \label{Eq:HSSmf}
 H^{\mathrm{mf}}_{S} &=& \sum_{\sigma} \left[Q^f
 \left(\langle \tilde{z}^\dagger_\sigma\rangle \tilde{z}_\sigma+ \langle \tilde{z}_\sigma\rangle \tilde{z}^\dagger_\sigma\right)
 + \lambda_\sigma S^z_{\sigma}\right] \nonumber\\
 &&+ \frac{U}{2} \left( \sum_{\sigma} S^z_{\sigma} \right)^2.
\end{eqnarray}

Eqs.~\eqref{Eq:Hfmf} and \eqref{Eq:HSSmf} represent the main formulation of the $U(1)$
 slave-spin approach at the mean-field level. Eq.~\eqref{Eq:HSSmf} consists of slave spins interacting at a single-site and can be easily solved. On the other hand, Eq.~\eqref{Eq:Hfmf} contains dispersive spinons.

The above mean-field Hamiltonians are solved self-consistently in the PM phase. The PM solution describes a metal to Mott insulator transition at finite value of $U$ when the system is at half-filling. Away from half-filling, it shows that the system is always metallic with a Fermi surface and finite quasiparticle spetral weight.

\subsection{Magnetic solutions of the $U(1)$ slave-spin theory for the Hubbard model}
We now consider the magnetic solution of the $U(1)$ slave-spin theory for the above single band Hubbard model, and we are in particular interested in the magnetic transition from PM state to magnetic ordered state. The key point is how to break the spin symmetry in above PM formula of Eqs.~\eqref{Eq:Hfmf} and \eqref{Eq:HSSmf}. Because the slave spin and fermionic spinon operators are decoupled at the mean-field level, one has to break the spin symmetry in both slave spin and spinon channels. Note that the slave-spin Hamiltonian includes only one single site, therefore, the breaking of spin symmetry can only via introducing an effective magnetic field coupled to the slave spin operator.

One na\"{\i}ve idea is to make use the spin dependent Lagrangian multiplier $\lambda_\sigma$. Since $\lambda_\sigma$ acts as an effective longitudinal field coupled to $S^z_{\sigma}$ in Eq.~\eqref{Eq:HSSmf}, the spin symmetry is then broken once $\lambda_{\uparrow}\neq\lambda_{\downarrow}$. Since $\lambda_\sigma$ appears in both Eqs.~\eqref{Eq:HSSmf} and \eqref{Eq:Hfmf}, the symmetry is also broken in the spinon channel once the constraint is satisfied. If we allow $\epsilon_k$ and $Q^f$ to be both spin dependent, and enforce $\epsilon_{k\uparrow}\neq\epsilon_{k\downarrow}$, $Q^f_{\uparrow}\neq Q^f_{\downarrow}$, then the above idea certainly works. But simply taking a spin dependent $\lambda_{\sigma}$ cannot break the spin symmetry spontaneously, namely, the system cannot evolve spontaneously from a PM state to a magnetic one. This is because $\lambda_{\sigma}$ is introduced to enforce the constraint in Eq.~\eqref{Eq:constraint}, and its contribution to the mean-field energy is $E_\lambda=\sum_{i\sigma}\lambda_{\sigma}(\langle S^z_{i\sigma} \rangle - \langle f^\dagger_{i\sigma} f_{i\sigma} \rangle + \frac{1}{2})$. Once can immediately show that $E_\lambda=0$ if the constraint is strictly satisfied. This means that there will be no energy gain in the magnetic state compared to the PM one, and therefore the magnetic solution can not be stabilized. similar argument applies to the idea of making the effective chemical potential $\tilde{\mu}_\uparrow\neq\tilde{\mu}_\downarrow$: the energy contribution of the $\tilde{\mu}_{\sigma}$ term, according to Eq.~\eqref{Eq:mutilde}, is also zero when the constraint is satisfied.

To make progress, we are inspired by the work of Ko and Lee~\cite{KoLee_PRB_2011} on investigating the magnetic properties of multiorbital Hubbard model in the slave rotor approach~\cite{FlorensGeorges}. They adopted a hybridized way in handling the interactions: taking into account the strong coupling effects of the Coulomb repulsion while performing a mean-field decomposition of the Hund's coupling term.

In our case of single band Hubbard model, we notice that the Coulomb repulsion can be rewritten as
\begin{equation}
 \frac{U}{2} \sum_{i,\sigma}(n_{i\sigma}-\frac{1}{2}) (n_{i\bar{\sigma}}-\frac{1}{2}) = -\frac{2}{3}U\sum_i \vec{\mathcal{S}}_i^2,
\end{equation}
where $\vec{\mathcal{S}}_i=\frac{1}{2} \sum_{\sigma\sigma^\prime} d^\dagger_{i\sigma}\vec{\tau}_{\sigma\sigma^\prime} d_{i\sigma^\prime}$ is the spin density operator, and $\vec{\tau}$ denote the Pauli matrices. We first consider the FM solution and assume $\mathcal{S}^z_i$ is ordered in the magnetic phase. We treat the XY and Z components of $\vec{\mathcal{S}}_i$ in a different way, namely, in symmetric and anti-symmetric channels, respectively,
\begin{eqnarray}
 (\mathcal{S}^x_i)^2+(\mathcal{S}^y_i)^2 && = -\frac{1}{2} (n_{i\uparrow} + n_{i\downarrow})^2 + (n_{i\uparrow} + n_{i\downarrow}) \\
 (\mathcal{S}^z_i)^2 && = \frac{1}{4} (n_{i\uparrow} - n_{i\downarrow})^2
\end{eqnarray}
Then we rewrite the interaction in the slave-spin representation,
\begin{eqnarray}
  && (n_{i\uparrow} + n_{i\downarrow})^2 \rightarrow [(\sum_\sigma S^z_{i\sigma})+1]^2, \\
  && (n_{i\uparrow} - n_{i\downarrow})^2 \rightarrow (S^z_{i\uparrow}-S^z_{i\downarrow})(f^\dagger_{i\uparrow}f_{i\uparrow} - f^\dagger_{i\downarrow}f_{i\downarrow}).
\end{eqnarray}
The Hamiltonian is then, up to a constant term,
\begin{eqnarray}
\label{Eq:HamSSMag}  H && = \frac{1}{2}\sum_{ij\sigma} t_{ij}
 \hat{z}^\dagger_{i\sigma} \hat{z}_{j\sigma} f^\dagger_{i\sigma} f_{j\sigma}
 -\mu \sum_{i\sigma}  f^\dagger_{i\sigma}
 f_{i\sigma}
 \nonumber\\
 && - \sum_{i\sigma} \lambda_{i\sigma}\left(f^\dagger_{i\sigma}
 f_{i\sigma}-S^z_{i\sigma}\right) + \frac{U}{3} \sum_i \left( \sum_{\sigma} S^z_{i\sigma} \right)^2 \nonumber\\
 && -\frac{U}{6} \sum_i \left(S^z_{i\uparrow}-S^z_{i\downarrow}\right) \left(f^\dagger_{i\uparrow}f_{i\uparrow} - f^\dagger_{i\downarrow}f_{i\downarrow}\right).
\end{eqnarray}
Defining the magnetic order parameter $m=\langle S^z_{i\uparrow}-S^z_{i\downarrow} \rangle = n^f_{i\uparrow} - n^f_{i\downarrow}$ and decompose the slave spin and fermionic spinon operators, we obtain the mean-field Hamiltonians in the two sectors for the FM phase
\begin{eqnarray}
 \label{Eq:Hfmfmag}
 H^{\mathrm{mf}}_f &=&  \sum_{k\sigma}\left[ \epsilon_{k} \langle \tilde{z}^\dagger_{\sigma} \rangle
  \langle \tilde{z}_{\sigma} \rangle + (\tilde{\mu}_\sigma-\lambda_\sigma-\mu)\right] f^\dagger_{k\sigma} f_{k\sigma},\nonumber\\
   && -\frac{U}{6}m\sum_{k\sigma} \sigma f^\dagger_{k\sigma} f_{k\sigma},\\
 \nonumber \\
 \label{Eq:HSSmfmag}
 H^{\mathrm{mf}}_{S} &=& \sum_{\sigma} \left[Q^f
 \left(\langle \tilde{z}^\dagger_\sigma\rangle \tilde{z}_\sigma+ \langle \tilde{z}_\sigma\rangle \tilde{z}^\dagger_\sigma\right)
 + \lambda_\sigma S^z_{\sigma}\right] \nonumber\\
 &&+ \frac{U}{3} \left( \sum_{\sigma} S^z_{\sigma} \right)^2 - \frac{U}{6}m \sum_{\sigma} \sigma S^z_{\sigma}.
\end{eqnarray}
In this formulation, a FM sate can be stabilized because the magnetic terms (the last term in Eq.~\eqref{Eq:Hfmfmag} and in Eq.~\eqref{Eq:HSSmfmag}) gain energy as indicated by the negative sign in front of them. The FM solution can then be obtained by diagonalizing the above Hamiltonians in a self-consistent way.

The AFM solution can be obtained in a similar way. As an example, we consider the N\'{e}el antiferromagnet on the square lattice. Still assuming $\mathcal{S}^z$ is ordered, then in the AFM state, we can inverse the spins on one sublattice by performing an $SU(2)$ spin rotation.  This sends $(d_{A\sigma}, d_{B\sigma})$ to $(d_{A\sigma}, d_{B\bar{\sigma}})$, where $A$ and $B$ are sublattice indices.
After the rotation, the n.n. hopping term in Eq.~\eqref{Eq:Ham_tot} becomes a hybridization between opposite spin flavors
\begin{equation}
 \sum_{\langle ij\rangle} t_{ij} d^\dagger_{i\sigma} d_{j\sigma} \rightarrow \sum_{\langle ij\rangle} t_{ij} d^\dagger_{i\sigma} d_{j\bar{\sigma}}.
\end{equation}
The AFM order then corresponds to a FM order in the rotated basis with $m$ now denotes the sublattice magnetization. One can then obtain the solution by replacing
\begin{eqnarray}
 \epsilon_{k} \langle \tilde{z}^\dagger_{\sigma} \rangle
  \langle \tilde{z}_{\sigma} \rangle && \rightarrow \epsilon_{k} \langle \tilde{z}^\dagger_{\sigma} \rangle
  \langle \tilde{z}_{\bar{\sigma}} \rangle \\
 Q^f
 \left(\langle \tilde{z}^\dagger_\sigma\rangle \tilde{z}_\sigma+ \langle \tilde{z}_\sigma\rangle \tilde{z}^\dagger_\sigma\right) && \rightarrow Q^f
 \left(\langle \tilde{z}^\dagger_\sigma\rangle \tilde{z}_{\bar{\sigma}}+ \langle \tilde{z}_{\bar{\sigma}}\rangle \tilde{z}^\dagger_\sigma\right)
\end{eqnarray}
in Eqs.~\eqref{Eq:Hfmfmag} and \eqref{Eq:HSSmfmag}, and diagonalizing the Hamiltonians self-consistently.

\section{Results}
\label{Sec:Results}
\begin{figure}[h!]
\centering\includegraphics[
width=70mm
]{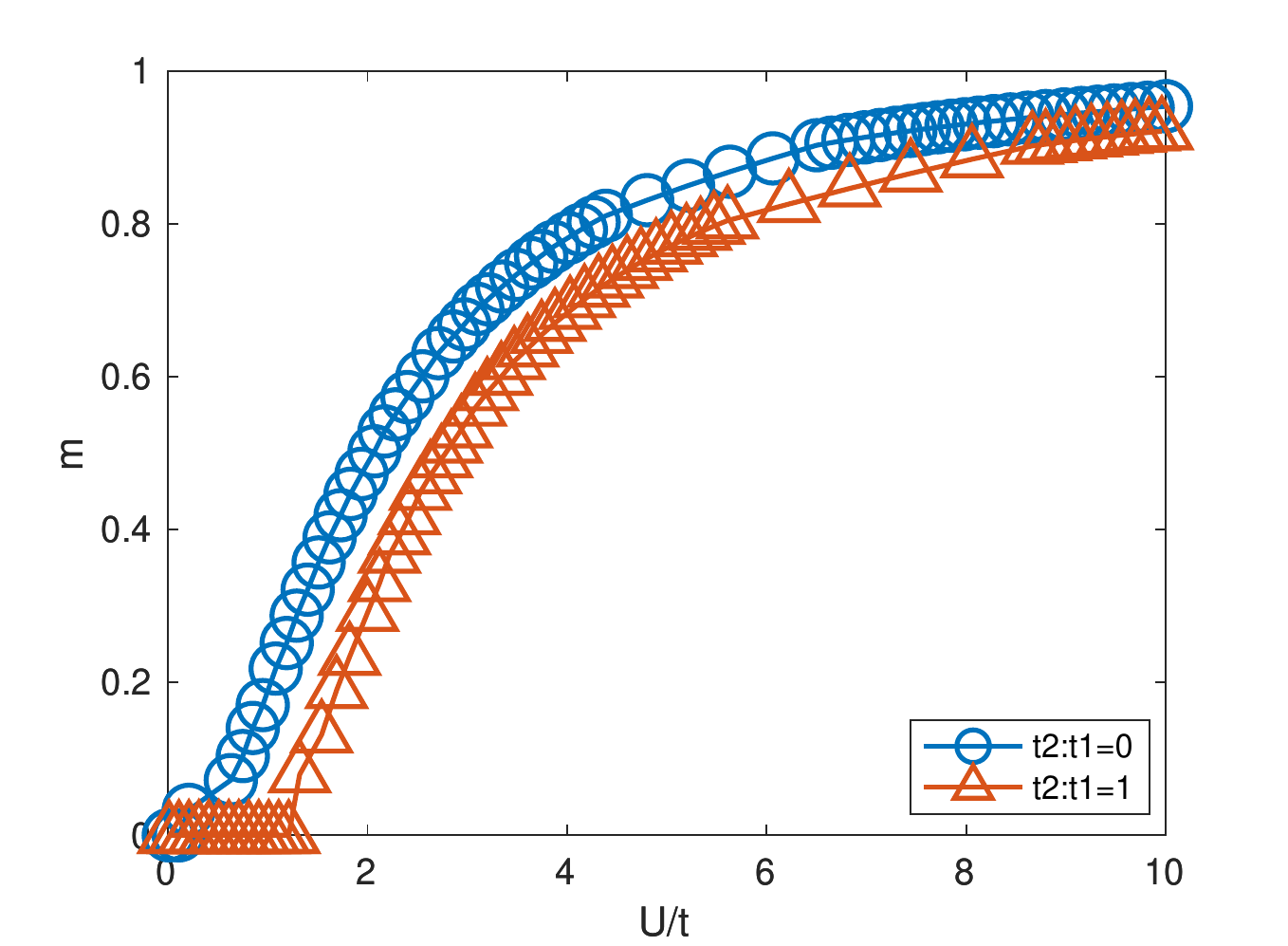}
\caption{(Color online) 
Evolution of the sublattice magnetization $m$ with $U/t$ of the Hubbard model at half-filling for $t_1=t$, $t_2=0$ and $t_1=t_2=t$, respectively. In either case, there is a PM to AFM transition above which $m>0$. 
}
\label{fig:1}
\end{figure}

Now we show some results about the FM and AFM states in the single band Hubbard model based on slave-spin approach described in Sec.~\ref{Sec:Model}. We first discuss results at half-filling. In the PM phase, it is expected that~\cite{Medici_PRB_2005,Yu_PRB_2012} a metal-to-Mott-insulator transition takes place at $U_c\sim 2D$ where $D$ is the bandwidth of the electrons in the noninteracting limit. For the model on square lattice with n.n. hopping $t_1=t$, $D=8t$.

This picture changes when considering AFM ordering. As shown in Fig.~\ref{fig:1}, a transition to AFM phase takes place at infinitesimal $U$ vale. At $U=D=8t$, the sublattice magnetization $m\approx0.9$, which is close to saturation. The result of $U_{\rm{AF}}\approx0$ originates from perfect nesting of the Fermi surface in the model, which causes gap opening under interaction. This scenario is captured by our mean-field approach. 

We then study the effect of a next n.n. hopping $t_2$ to the AFM ordering by taking $t_1=t_2=t$. Including a finite $t_2$ disturbs the perfect nesting condition, and the AFM transition takes place at a larger value $U_{\rm{AF}}=1.4t$ (Fig.~\ref{fig:1}), and for same $U$, the sublattice magnetization is smaller than that in the $t_2=0$ case. 

\begin{figure}[h!]
\centering\includegraphics[
width=70mm
]{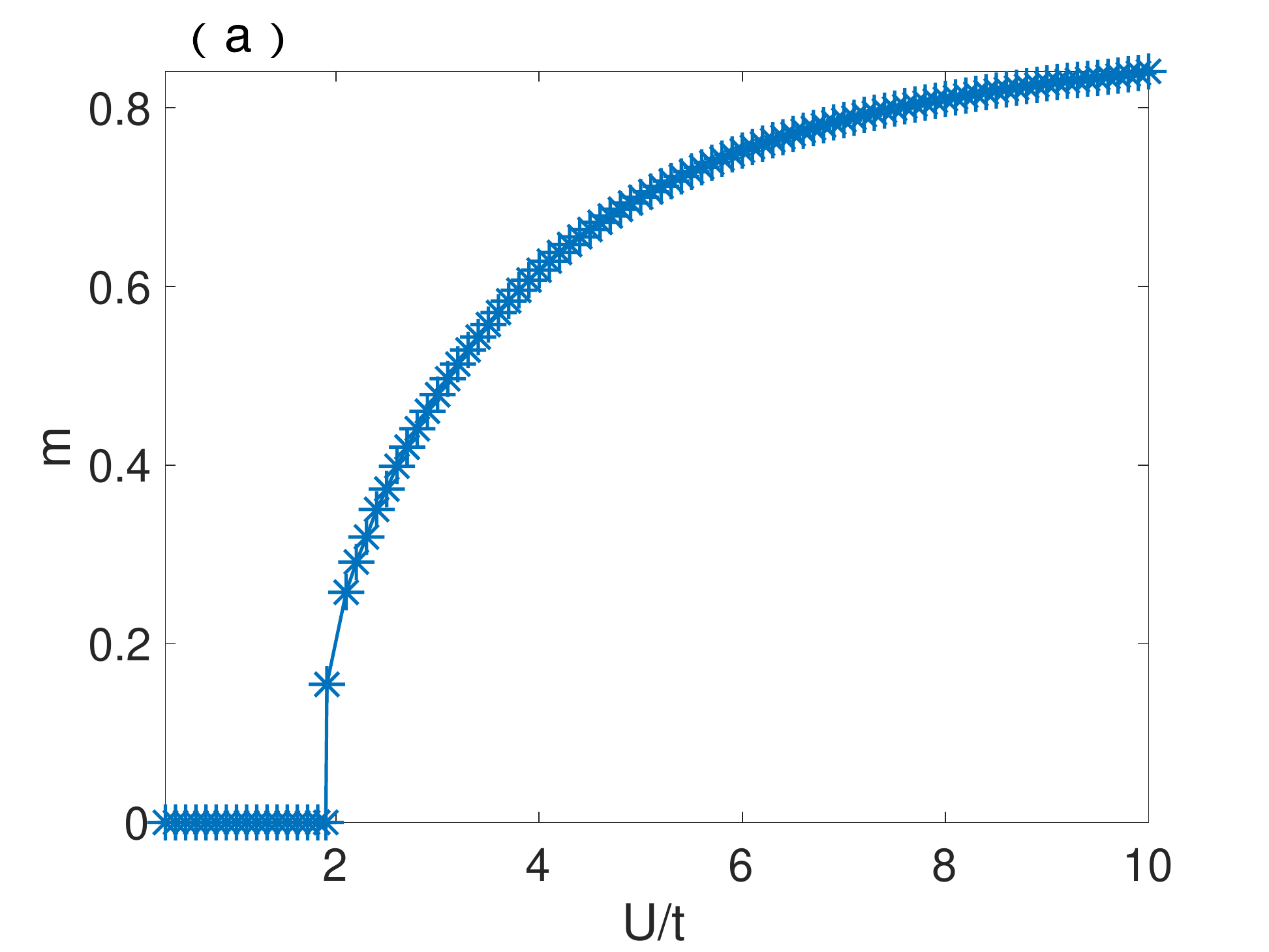}
\centering\includegraphics[
width=70mm
]{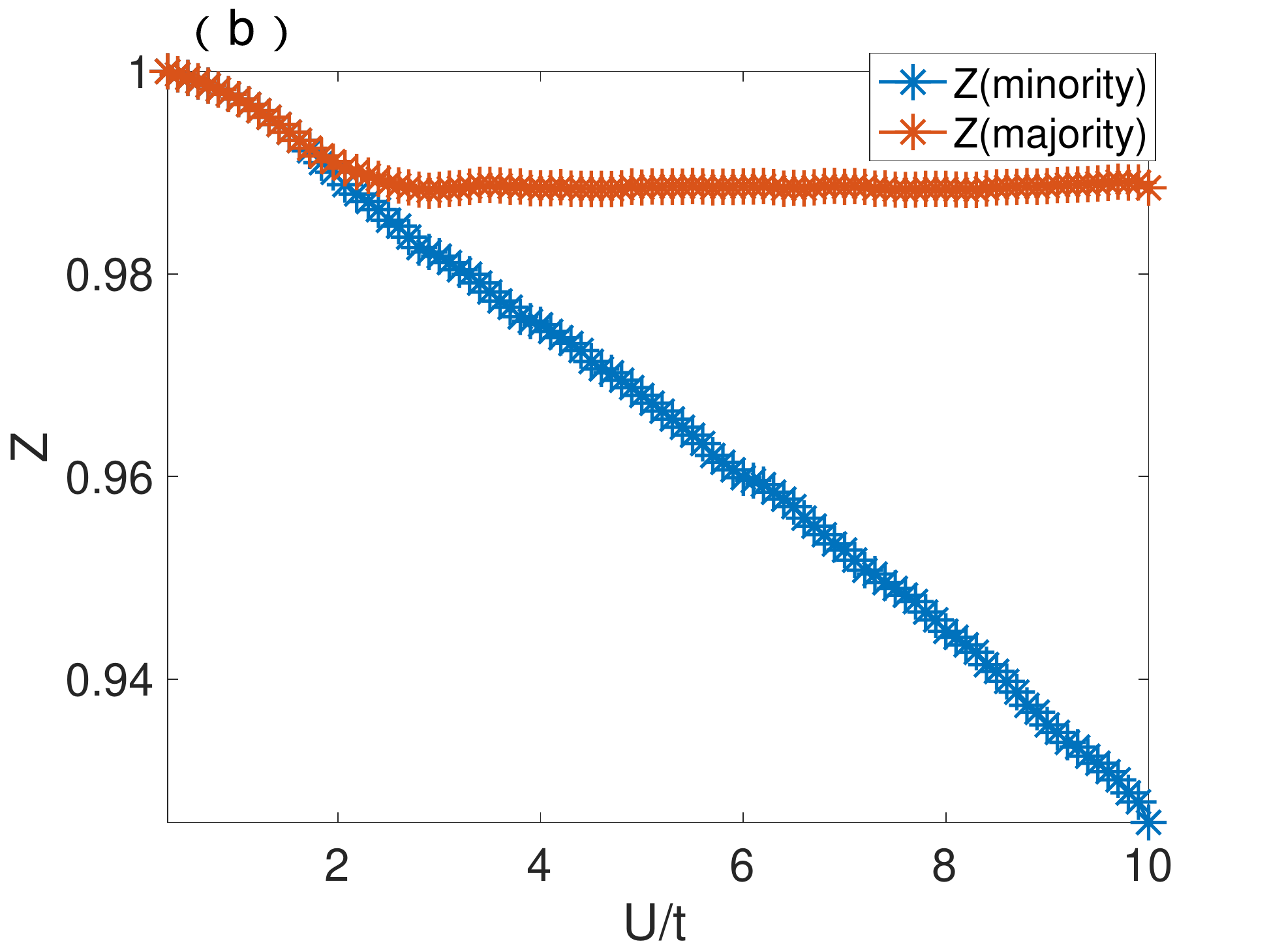}
\caption{(Color online) (a): Evolution of the sublattice magnetization $m$ with $U/t$ of the Hubbard model at $t_1=t$, $t_2=0$ and $n=1.1$, showing a PM to AFM transition. (b): Corresponding evolution of quasiparticle spectral weights.
}
\label{fig:2}
\end{figure}

We now turn to magnetic properties at finite doping, with $t_1=t$ and $t_2=0$. In Fig.~\ref{fig:2} we show the sublattice magnetization and quasiparticle spectral weights at $n=1.1$, where $n=\sum_{\sigma} \langle n_\sigma \rangle$. We find a PM to AFM transition at $U_{\rm{AF}}\approx 2t$. In the AFM phase, the magnetization increases with $U$ rapidly near the transition point, implying a first-order transition. For about $m\gtrsim0.2$ it gradually increases with $U$, and approaches saturation value $m=0.9$ very slowly at large $U$ values. In the AFM phase, quasiparticle spectral weights of the majority and minority spins evolve differently with $U$. $Z$ of electrons with majority spin reduces very slowly with $U$ and approaches a constant value close to $1$, whereas $Z$ of electrons with minority spin drops much faster and almost linearly with $U$ within the range of our calculation. The separation of $Z$ for different spin flavors reflects asymmetric electron fillings. With increasing $U$, the filling number of electrons with majority spin increases quickly and approaches close to $1$ (fully occupied), and these electrons are less affected by the correlation effect than those with minority spin, whose filling number approaches a nonzero number ($n\sim0.1$) and thus feel stronger band renormalization effect.

\begin{figure}[h!]
\centering\includegraphics[
width=70mm
]{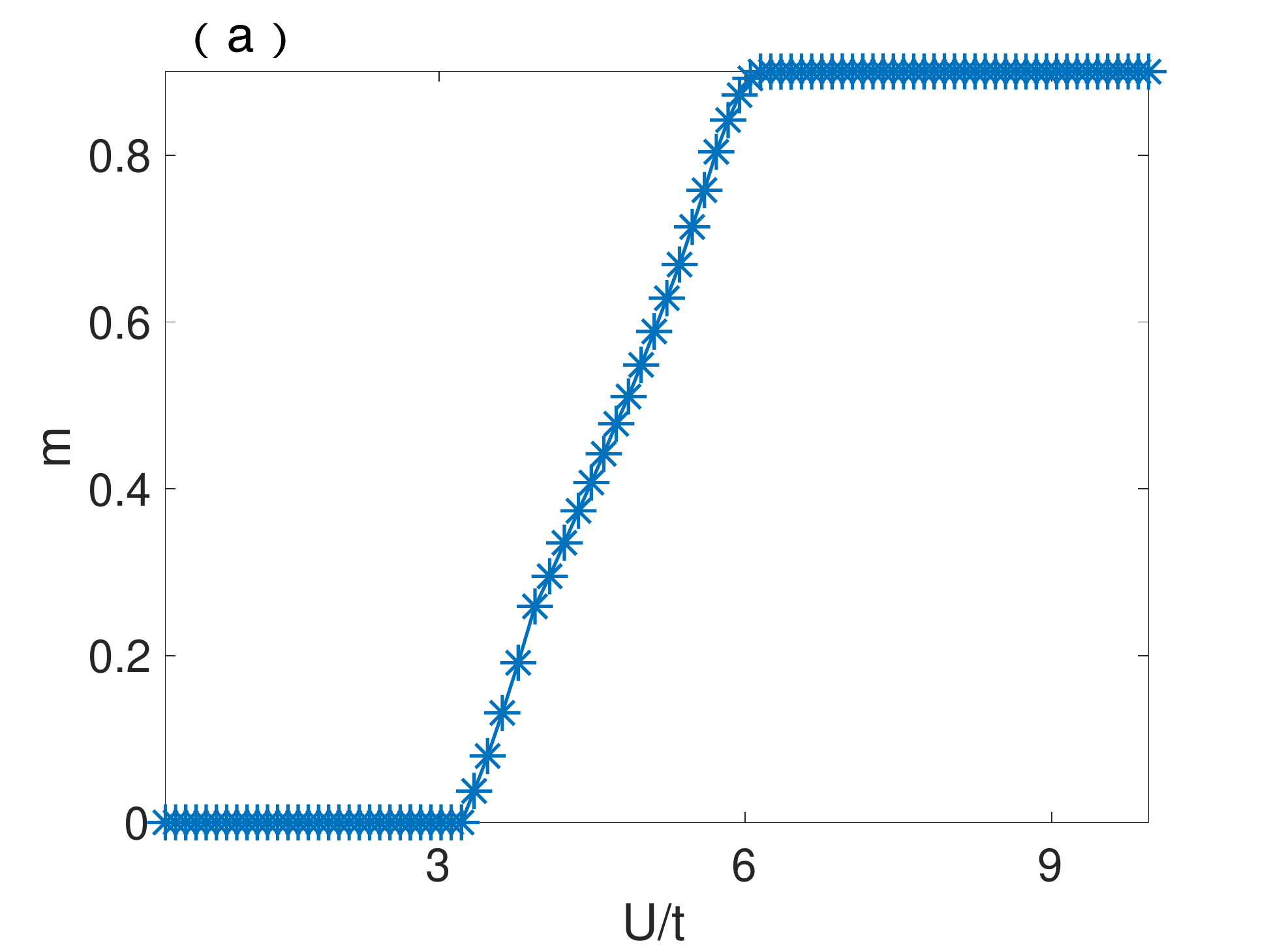}
\centering\includegraphics[
width=70mm
]{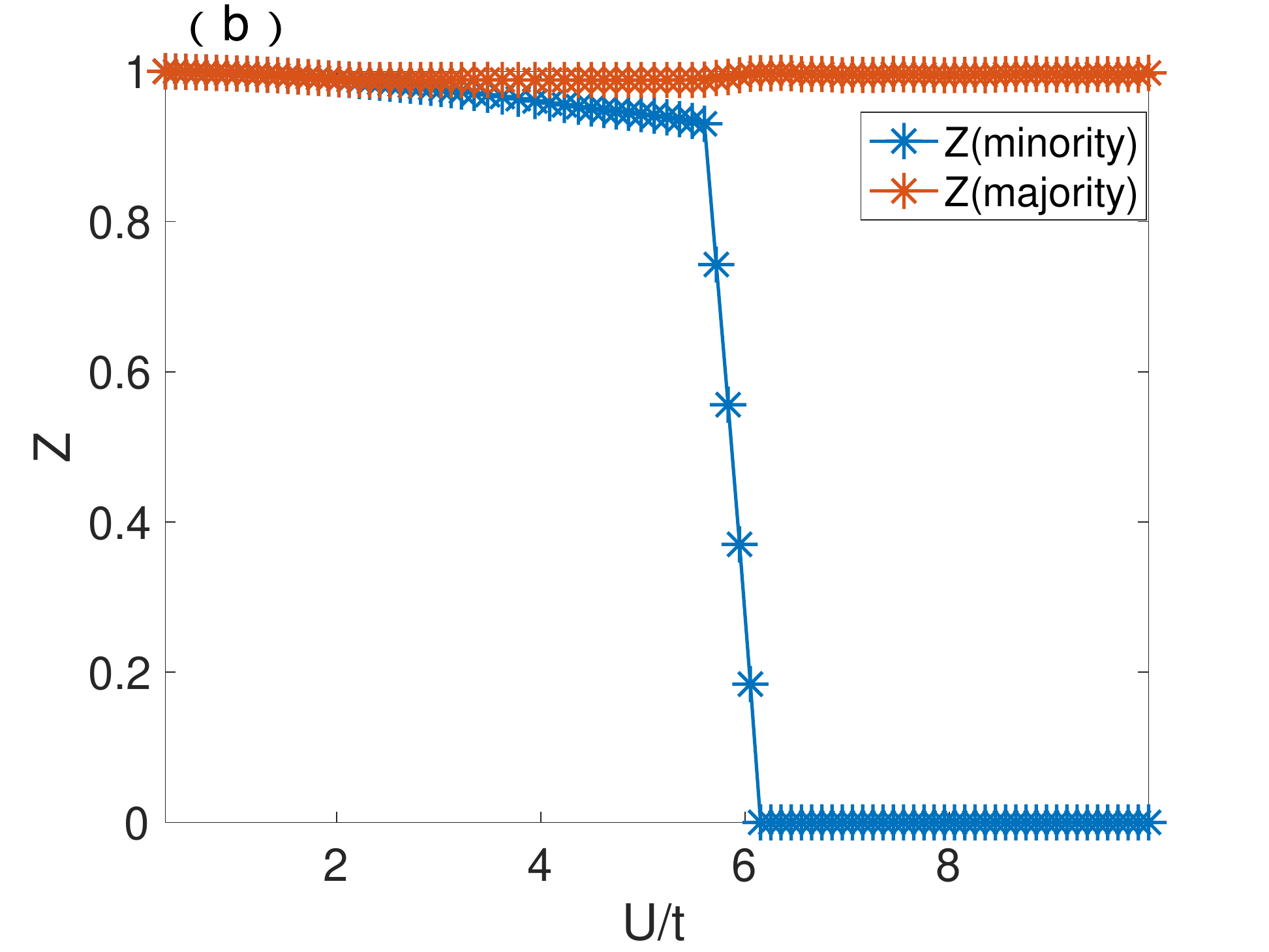}
\caption{(Color online) (a): Evolution of the uniform magnetization $m$ with $U/t$ of the Hubbard model at $t_1=t$, $t_2=0$ and $n=1.1$, showing a PM to FM transition. (b): Corresponding evolution of quasiparticle spectral weights.
}
\label{fig:3}
\end{figure}

Besides the AFM state, a FM state can also be stabilized at finite doping. As shown in Fig.~\ref{fig:3}(a), a continuous PM to FM transition takes place at $U_{\rm{FM}}\approx3.5t$, the uniform magnetization $m$ increases rapidly and saturates at $U_{\rm{s}}\approx6t$. In the FM state, the quasiparticle spectral weights of the majority (up) and minority (down) spins separate, similar to the AFM case. $Z$ of electrons with the majority spin slightly decreases, then increases to $1$ when $m$ saturates, reflecting that the band width of a fully occupied band is not renormalized by the electron correlations. On the other hand, $Z$ of electrons with the minority spin decreases rapidly with $U$ and vanishes when $m$ saturates. This can be understood as follows. In the FM phase a gap is not immediately open at the Fermi level, and both majority and minority electrons are itinerant. Because the density of the minority electrons approaches a fractional number, the band is strongly renormalized. Once $m$ saturates, the filling number is fixed, and the only way it gains energy is to shrink the band width to zero. This is similar to a Mott insulator in which the quasiparticle weight is zero and the spinon has a finite Fermi surface. Therefore, the FM state above $U_{\rm{s}}$ can be viewed as a spin-selective Mott phase, where the spin up electrons are itinerant and the spin down electrons are localized.

\begin{figure}[h!]
\centering\includegraphics[
width=70mm
]{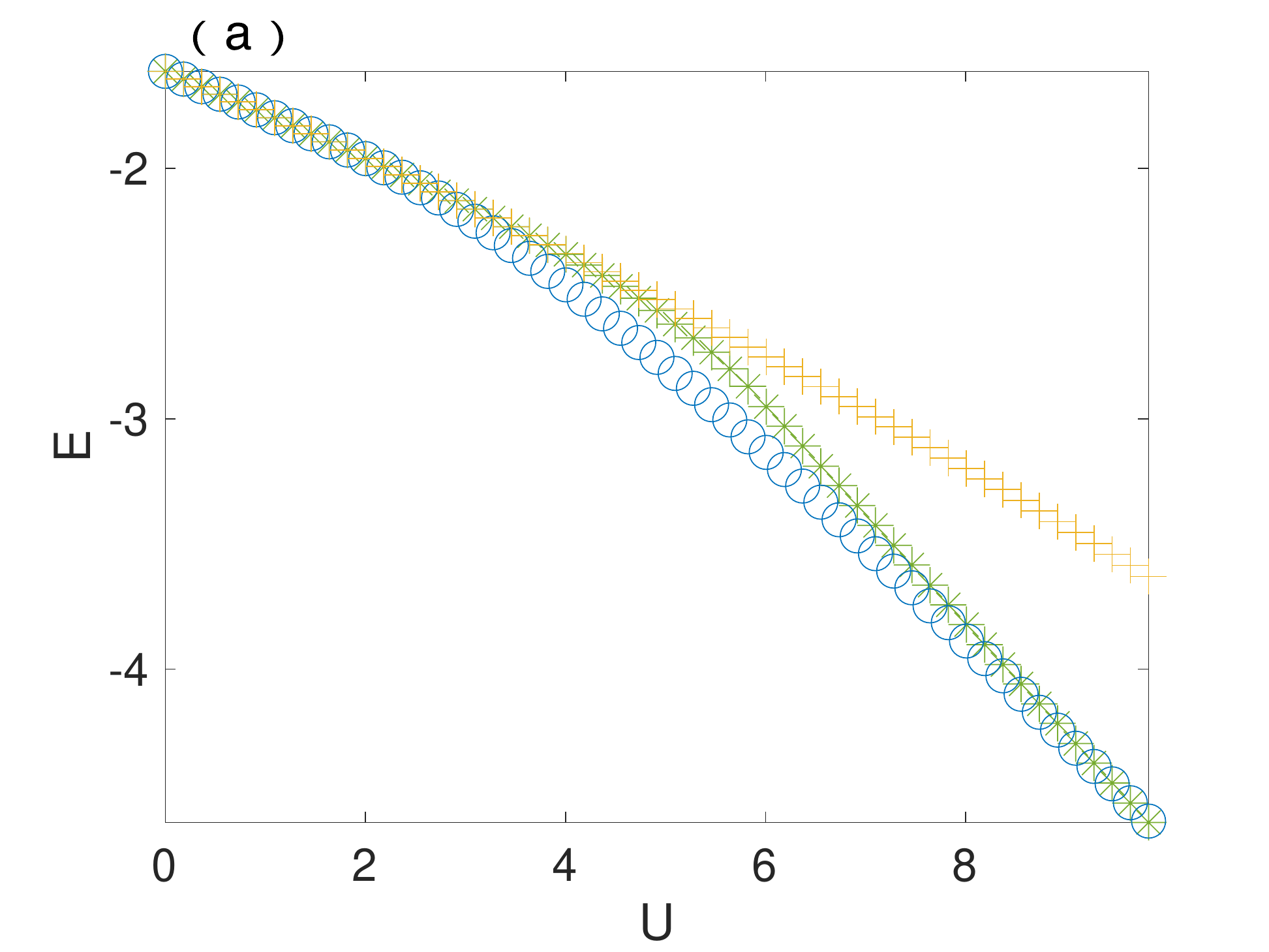}
\centering\includegraphics[
width=70mm
]{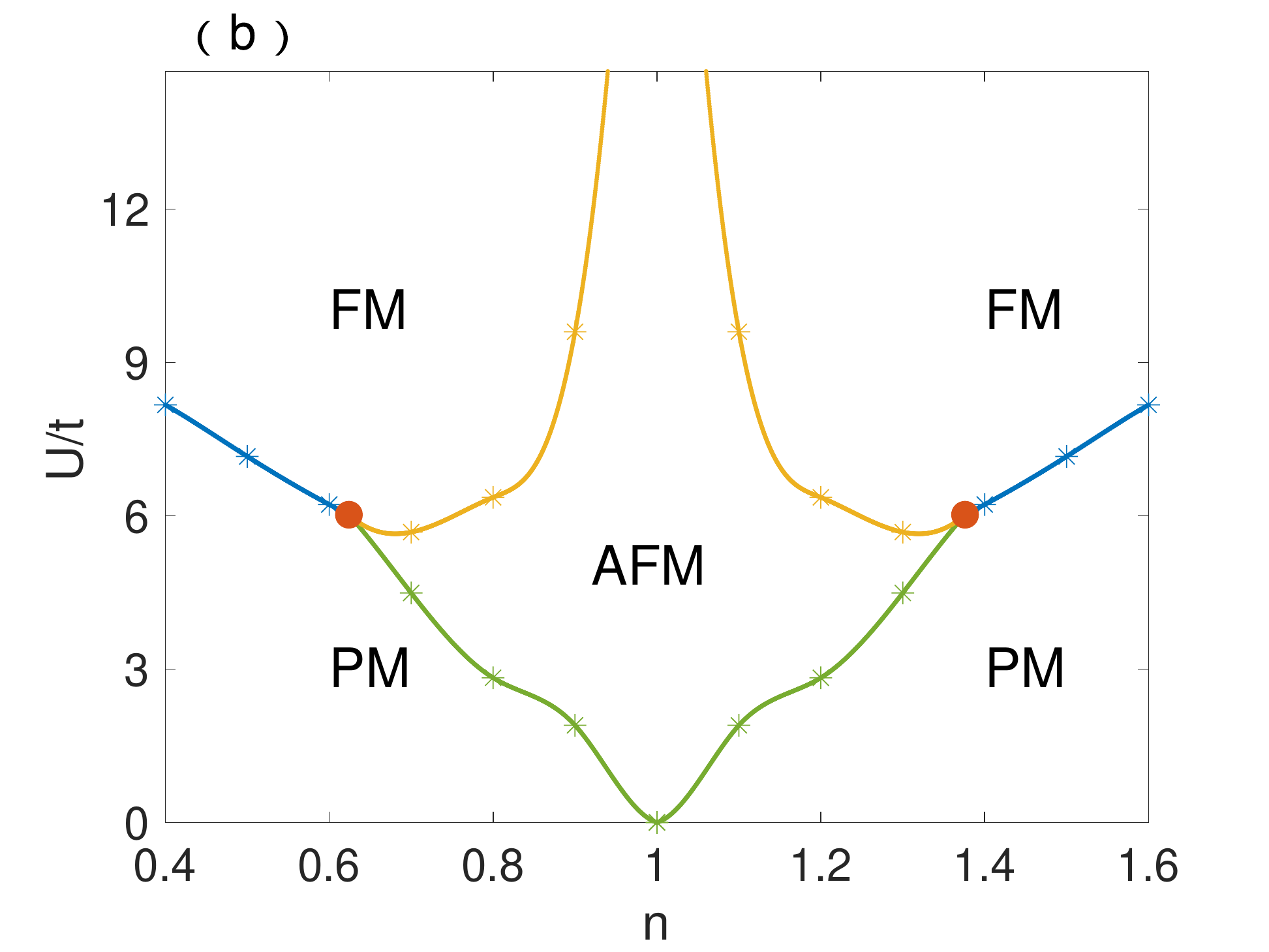}
\caption{(Color online) (a): Comparison of ground-state energies of the paramagnetic (PM), ferromagnetic (FM), and antiferromagnetic (AFM) solutions of the Hubbard model at $n=1.1$ in the slave-spin approach. The ground state at each $U$ value is determined to be the state with the lowest energy. (b): Ground-state phase diagram in the $n$-$U/t$ plane. The red dot is a triple point where PM, FM, and AFM phases meet. The phase diagram is symmetric about $n=1$, reflecting the particle-hole symmetry of the model.
}
\label{fig:4}
\end{figure}

With both FM and AFM states stabilized at a fixed filling, we can then determine the exact ground state by comparing their energies. Fig.~\ref{fig:4}(a) shows the energies of the PM, FM, and AFM solutions of the Hubbard model at filling number $n=1.1$ within the slave-spin theory. It is clearly seen that transition from PM to magnetic ordered states lowering the energy. With increasing $U$ the system first undergoes a PM to AFM transition at $U_{\rm{AF}}$. Then at sufficiently large $U$ the FM state with saturated magnetization can have a lower energy than the AFM state whose sublattice magnetization is not completely saturated, and the system eventually enters in a FM state.

By comparing the energies we obtain a ground-state phase diagram of the Hubbard model for generic filling factor. The AFM phase is stabilized in the intermediate to strong $U/t$ regime around half-filling ($n=1$) up to a triple point near $n=1.4$ (and $n=0.6$) where the PM, FM, and AFM phases meet. For doping above this point, there is a direct PM to FM transition.

\section{Discussions and Conclusion}
\label{Sec:Discussions}

In the AFM phase, the $Z$ value for either spin flavor is larger than that in the PM state, and strong electron correlation effects present when $U\gg D$. This is because the AFM order pushes electrons with both spins away from half-filling. Surprisingly, electrons in the FM state exhibit strong spin dependent correlation effects: the spin up band is less renormalized whereas the spin down one is strongly renormalized and its band width shrinks to zero when the magnetization saturates. In the phase diagram of Fig.~\ref{fig:4}(b), the larger $U$ regime away from half-filling is dominant by the FM state with saturated magnetization. Whether such a state survives fluctuations beyond the mean-field level and whether this state is connected to the Nagaoka ferromagnetism~\cite{Nagaoka_PR_1966} in the infinite $U$ limit deserve further investigation.

Our slave-spin approach to the magnetic solutions is very efficient and can be easily generalized to investigate magnetic properties of multiorbital Hubbard models. Though magnetic ground states can be stabilized within the present theory, correlation effects are underestimated within the single-site approximation we adopted. This issue can be relieved by generalizing the method to a cluster slave-spin approach~\cite{Lee_PRB_2017}.

In conclusion, we have developed a slave spin theory of the magnetic states in the single band Hubbard model. By decomposing the on-site Coulomb interaction in the magnetic channel in both the slave spin and the fermionic spinon sectors, we can stabilize both ferromagnetic and antiferromagnetic states with renormalized quasiparticle spectral weight at generic filling. The result is summarized in a ground-state phase diagram involving paramagnetic, ferromagnetic, and antiferromagnetic phases. While the antiferromagnetic state is stabilized in the intermediate and strong interaction regime near half-filling, the system is dominated by a strong coupling ferromagnetic state far from half-filling.

\acknowledgements
We
are grateful to E. Bascones, W. Ding, L. de' Medici, A. H. Nevidomskyy, and Q. Si,
for useful discussions.
This work has been supported by
the National Science Foundation of China Grant number 12174441.
R.Y. acknowledges the hospitality of Kavli Institute for Theoretical Physics during the follow-up discussions of the 2014 program on ``Magnetism, Bad Metals and Superconductivity: Iron Pnictides and Beyond'', where part of this work was discussed.





\begin{thebibliography}{99}

\bibitem{Dagotto_RMP_1994} E. Dagotto, ``Correlated electrons in high-temperature superconductors'', Rev. Mod. Phys. \textbf{66}, 763 (1994).

\bibitem{Lee_RMP_2006} P. A. Lee, N. Nagaosa, and X.-G. Wen, ``Doping a Mott insulator: Physics of high-temperature superconductivity'', Rev. Mod. Phys. \textbf{78}, 17 (2006).

\bibitem{Dai_RMP_2015} P. Dai, ``Antiferromagnetic order and spin dynamics in iron-based Superconductors'',
Rev. Mod. Phys. \textbf{87}, 855-896 (2015).

\bibitem{Si_NRM_2016} Q. Si, R. Yu and E. Abrahams,
``High Temperature Superconductivity in Iron Pnictides and Chalcogenides'', Nat. Rev. Mater. \textbf{1}, 16017 (2016).

\bibitem{Hirschfeld_CRP_2016} P. J. Hirschfeld, ``Gap Symmetry and Structure to Reveal the Pairing Mechanism in Fe-based Superconductors'',
Comptes Rendus Physique \textbf{17}, 197 (2016).

\bibitem{Wang_Science_2011} F. Wang and D.-H. Lee, ``The Electron-Pairing Mechanism of Iron-Based Superconductors'',
Science {\bf 332}, 200-204 (2011).

\bibitem{Imada_RMP_1998} M. Imada, A. Fujimori, and Y. Tokura, ``Metal-insulator transitions'', Rev. Mod. Phys. \textbf{70}, 1039 (1998).

\bibitem{Stewart_RMP_2001} G. R. Stewart, ``Non-Fermi-liquid behavior in $d$- and $f$-electron metals'' Rev. Mod. Phys. \textbf{73}, 797 (2001).

\bibitem{Fradkin_RMP_2015} E. Fradkin, S. A. Kivelson, and J. M. Tranquada, ``Theory of intertwined orders in high temperature superconductors'', Rev. Mod. Phys. \textbf{87}, 457 (2015).

\bibitem{PaschenSi_NRP_2021} S. Paschen and Q. Si, ``Quantum phases driven by strong correlations'', Nat. Rev. Phys. \textbf{3}, 9-26 (2021).

\bibitem{Arovas_ARCMP_2022} D. P. Arovas, E. Berg, S. Kivelson, and S. Raghu, ``The Hubbard Model'', Ann. Rev. Condens. Matt. Phys. \textbf{13}, 239 (2022).

\bibitem{Medici_PRB_2005} L. de' Medici, A. Georges, and S. Biermann, ``Orbital-selective Mott transition in multiband systems: Slave-spin representation and dynamical mean-field theory'', Phys. Rev. B \textbf{72}, 205124 (2005).

\bibitem{Medici_PRL_2009}
L. de' Medici,  S. R. Hassan,  M. Capone, and X. Dai,
``Orbital-Selective Mott Transition out of Band Degeneracy Lifting'',
Phys. Rev. Lett. \textbf{102},126401 (2009)

\bibitem{Yu_PRB_2011}
R. Yu and Q. Si,
``Mott Transition in Multiorbital Models for Iron Pnictides'',
Phys. Rev. B {\bf 84}, 235115 (2011).

\bibitem{Medici_PRB_2011}
L. de' Medici,
``Hund's coupling key role in multi-orbital correlations",
Phys. Rev. B {\bf 83}, 205112 (2011).

\bibitem{Yu_PRB_2012}
R. Yu and Q. Si,
``$U(1)$ Slave-spin theory and its application to Mott transition in a multiorbital model for iron pnictides'',
Phys. Rev. B{\bf 86}, 085104 (2012).

\bibitem{Yu_PRL_2013}
R. Yu and Q. Si,
``Orbital-selective Mott Phase in Multiorbital Models for Alkaline Iron
Selenides K$_{\rm 1-x}$Fe$_{\rm 2-y}$Se$_{\rm 2}$'',
Phys.~Rev.~Lett. {\bf 110}, 146402 (2013).

\bibitem{Medici_PRL_2014}
L. de' Medici, G. Giovannetti, and M. Capone,
``Selective Mott Physics as a Key to Iron Superconductors'',
Phys. Rev. Lett. 112, 177001 (2014).

\bibitem{Fanfarillo_PRB_2015} L. Fanfarillo and E. Bascones, ``Electronic correlations in Hund metals'', Phys. Rev. B \textbf{92}, 075136 (2015).

\bibitem{Yu_PRB_2017} R. Yu and Q. Si, ``Orbital-selective Mott phase in multiorbital models for iron pnictides and chalcogenides'', Phys. Rev. B \textbf{96}, 125110 (2017).

\bibitem{Komijani_PRB_2017} Y. Komijani and G. Kotliar, ``Analytical slave-spin mean-field approach to orbital selective Mott insulators", Phys. Rev. B \textbf{96}, 125111 (2017).

\bibitem{Yu_PRL_2018} R. Yu, J.-X. Zhu, and Q. Si, ``Orbital Selectivity Enhanced by Nematic Order in FeSe'', Phys. Rev. Lett. \textbf{121}, 227003 (2018).

\bibitem{Yang_CPB_2019} W.-W. Yang, H.-G. Luo, and Y. Zhong, ``Benchmarking the simplest slave-particle theory with Hubbard dimer'', Chin. Phys. B \textbf{28}, 107103 (2019).

\bibitem{Pizarro_JPC_2019} J. M. Pizarro, M. J. Calder\'{o}n, and E. Bascones, ``The nature of correlations in the insulating states of twisted bilayer graphene'', J. Phys. Commun. \textbf{3}, 035024 (2019).

\bibitem{Arribi_PRB_2021} P. V. Arribi and L. de' Medici, ``Hund's metal crossover and superconductivity in the 111 family of iron-based superconductors'', Phys. Rev. B \textbf{104}, 125130 (2021).

\bibitem{Georges_RMP_1996} A. Georges, G. Kotliar, W. Krauth, and M. J. Rozenberg, ``Dynamical mean-field theory of strongly correlated fermion systems and the limit of infinite dimensions'', Rev. Mod. Phys. \textbf{68}, 13 (1996).

\bibitem{Kotliar_RMP_2006} G. Kotliar, S. Y. Savrasov, K. Haule, V. S. Oudovenko, O. Parcollet, and C. A. Marianetti, ``Electronic structure calculations with dynamical mean-field theory'', Rev. Mod. Phys. \textbf{78}, 865 (2006).

\bibitem{Nandkishore_PRB_2012}
R. Nandkishore, M. A. Metlitski, and T. Senthil,
``Orthogonal Metals: The simplest non-Fermi liquids''
Phys. Rev. B {\bf 86}, 045128 (2012).

\bibitem{KotliarRuckenstein}
G. Kotliar and A. Ruckenstein,
``A new functional integral approach to strongly correlated Fermi systems: the Gutzwiller approximation as a saddle point",
Phys. Rev. Lett. \textbf{57}, 1362 (1986).

\bibitem{KoLee_PRB_2011} W.-H. Ko and P. A. Lee, ``Magnetism and Mott transition: A slave-rotor study'', Phys. Rev. B \textbf{83}, 134515 (2011).

\bibitem{FlorensGeorges}
S. Florens, A. Georges,
``Slave-rotor mean-field theories of strongly correlated systems and the Mott transition in finite dimensions'',
Phys. Rev. B {\bf 70} 035114 (2004).

\bibitem{Nagaoka_PR_1966} Y. Nagaoka, ``Ferromagnetism in a Narrow, Almost Half-Filled $s$ Band'', Phys. Rev. \textbf{147}, 392 (1966).

\bibitem{Lee_PRB_2017} W.-C. Lee and T.-K. Lee, ``Antiferromagnetism in the Hubbard model using a cluster slave-spin method'', Phys. Rev. B \textbf{96}, 115114 (2017).

\end{thebibliography}
\end{document}